\documentclass[12pt]{article}
\usepackage[utf8]{inputenc}
\usepackage[T1]{fontenc}
\usepackage{subcaption}
\usepackage{caption}
\usepackage{cite}
\makeatletter
\def\@cite#1#2{\textsuperscript{[#1\if@tempswa , #2\fi]}}
\makeatother
\usepackage{array,graphicx,dcolumn,multirow,hevea,abstract,hanging,fancyhdr,float}
\usepackage{footnote}
\usepackage{algorithm}
\usepackage{algpseudocode}
\usepackage[many]{tcolorbox}  

\tcbset{
    sharp corners,
    colback = white,
    before skip = 0.2cm,    
    after skip = 0.5cm      
}         

\newtcolorbox[auto counter, number within=section]{boxA}[2][]{%
    fontupper = \small,
    boxrule = 1.5pt,
    colframe = black, 
    title=Algorithm~\thetcbcounter: #2,
    #1
}

\newtcolorbox{boxB}{
    fontupper = \bf\color{main}, 
    boxrule = 1.5pt,
    colframe = main,
    rounded corners,
    arc = 5pt   
}

\newtcolorbox{boxC}{
    colback = sub, 
    boxrule = 0pt  
}

\newtcolorbox{boxD}{
    colback = sub, 
    colframe = main, 
    boxrule = 0pt, 
    toprule = 3pt, 
    bottomrule = 3pt 
}

\newtcolorbox{boxE}{
    enhanced, 
    boxrule = 0pt, 
    borderline = {0.75pt}{0pt}{main}, 
    borderline = {0.75pt}{2pt}{sub} 
}

\newtcolorbox{boxF}{
    colback = sub,
    enhanced,
    boxrule = 1.5pt, 
    colframe = white, 
    borderline = {1.5pt}{0pt}{main, dashed} 
}

\newtcolorbox{boxG}{
    enhanced,
    boxrule = 0pt,
    colback = sub,
    borderline west = {1pt}{0pt}{main}, 
    borderline west = {0.75pt}{2pt}{main}, 
    borderline east = {1pt}{0pt}{main}, 
    borderline east = {0.75pt}{2pt}{main}
}

\newtcolorbox{boxH}{
    colback = sub, 
    colframe = main, 
    boxrule = 0pt, 
    leftrule = 6pt 
}

\newtcolorbox{boxI}{
    colback = sub, 
    colframe = main, 
    boxrule = 0pt, 
    toprule = 6pt 
}

\newtcolorbox{boxJ}{
    sharpish corners, 
    colback = sub, 
    colframe = main, 
    boxrule = 0pt, 
    toprule = 4.5pt, 
    enhanced,
    fuzzy shadow = {0pt}{-2pt}{-0.5pt}{0.5pt}{black!35} 
}

\newtcolorbox{boxK}{
    sharpish corners, 
    boxrule = 0pt,
    toprule = 4.5pt, 
    enhanced,
    fuzzy shadow = {0pt}{-2pt}{-0.5pt}{0.5pt}{black!35} 
}

\newtcolorbox{boxL}{
    fontupper = \color{main},
    rounded corners,
    arc = 6pt,
    colback = sub, 
    colframe = main!50, 
    boxrule = 0pt, 
    bottomrule = 4.5pt 
}

\newtcolorbox{boxM}{
    fontupper = \color{white},
    rounded corners,
    arc = 6pt,
    colback = main!80, 
    colframe = main, 
    boxrule = 0pt, 
    bottomrule = 4.5pt,
    enhanced,
    fuzzy shadow = {0pt}{-3pt}{-0.5pt}{0.5pt}{black!35}
}
\usepackage{amsmath} 
\usepackage[labelfont=sc,textfont=sf]{caption}
\usepackage[hyperfootnotes=false,breaklinks=true]{hyperref} 
\usepackage{booktabs} 
\newcolumntype{L}[1]{>{\raggedright\arraybackslash }p{#1}} 
\newcolumntype{C}[1]{>{\centering\arraybackslash }p{#1}}
\newcolumntype{R}[1]{>{\raggedleft\arraybackslash }p{#1}}
\newcolumntype{d}[1]{D{.}{.}{#1}} 

\let\tempone\itemize
\let\temptwo\enditemize
\let\tempthree\enumerate
\let\tempfour\endenumerate
\renewenvironment{itemize}{\tempone\setlength{\itemsep}{0pt}}{\temptwo}
\renewenvironment{enumerate}{\tempthree\setlength{\itemsep}{0pt}}{\tempfour}

\title{Single-Sequence-Based Protein Secondary Structure Prediction using One-Hot and Chemical Encodings of Amino Acids}

\author{
Hoa Trinh\thanks{Department of Physics, University of
  Southern California, USA}
  \and
  Satish Kumar Thittamaranahalli\thanks{Information Science Institute, USA}
}

\date{} 

\begin{document} 
\captionsetup[figure]{labelsep=space}
\begin{htmlonly}
\href{\jref}{\jhead}, \jdate, pp.\
\end{htmlonly}

\maketitle

\begin{abstract}
In protein secondary structure prediction, each amino acid in sequence is typically treated as a distinct category and represented by a one-hot vector. In this study, we developed two novel chemical representations for amino acids utilizing molecular fingerprints and the dimensionality reduction algorithm FastMap. We demonstrate that the two new chemical encodings can provide additional information about the interactions of amino acids in sequences that an LSTM-based model cannot capture with one-hot encoding alone. Compared to the latest LSTM-based model used in the single-sequence-based method SPOT-1D-Single, our ensemble model utilizing one-hot and chemical encodings achieves better accuracy across most test sets while requiring approximately nine times fewer trainable parameters for each encoding model. Our single-sequence-based method is valuable for its simplicity, lower resource requirements, and independence from external sequence data. It is beneficial when quick or preliminary predictions are needed or when data on homologous sequences is scarce.

\end{abstract}


\setlength{\baselineskip}{16pt plus.2pt}

\section{INTRODUCTION}
Protein secondary structure prediction is vital in biology, as knowing protein structures can provide crucial insights into their function. To date, deep learning-based prediction approaches can be categorized into three primary categories: profile-based approaches, protein language model approaches, and single-sequence-based approaches. 

The profile-based approaches extract evolutionary information from multiple sequence alignments of homologous sequences by employing techniques such as the Position-Specific Scoring Matrix (PSSM) profile\cite{bib9} or Hidden Markov Model (HMM) profile\cite{bib10}. MUFOLD-SS\cite{mufold}, a deep inception network, employs PSSM and HMM profiles as inputs. It achieves $86.5\%$ Q3 accuracy. 

Protein language model approaches generate embeddings for protein sequences by leveraging transformer-based models trained on extensive databases. SPOT-1D-LM\cite{1dlm} employs the embeddings generated by ProtTrans\cite{trans} and ESM-1b\cite{esm} as input. It achieves profile-based accuracy for various test sets. At the time of prediction, although the protein language model approach does not rely on multiple-sequence alignments, it utilizes pre-trained embeddings developed from 50 million protein sequences of the Uniref50 database. Therefore, SPOT-1D-LM, in a stricter sense, is not a single-sequence-based method.

Unlike profile-based and protein language model approaches, single-sequence-based methods do not use information from external databases. These methods rely solely on the information contained within a single sequence. Even though single-sequence-based methods often underperform compared to the two counterpart approaches, they generally have faster inference time due to no computational overhead associated with multiple-sequence alignment and embedding generation. They offer a significant advantage over multiple-sequence-alignment-based approaches for proteins with limited homologs\cite{bib15, bib12}. Most importantly, single-sequence-based methods could help us to understand the underlying mechanisms of secondary structure formation.

Single-sequence-based methods explain the formation of protein secondary
structures as a result of interactions between amino acids in sequence. Amino acid encodings are important for modeling their interactions. Since the twenty types of amino acids differ only by their side chains, it is encouraging to encode each amino acid using their side chain properties. Fauch{\`e}re et al. derived fifteen physicochemical parameters for the twenty side chains through experimental methods\cite{properties}. Building on this, Meiler et al. used seven of these parameters—including hydrophobicity, steric properties, and electronic characteristics—to encode each amino acid, achieving a $Q_{3}$ accuracy of 67\%. Recent single-sequence-based methods such as SPIDER3-Single\cite{bib12}, ProteinUnet\cite{bib17}, and SPOT-1D-Single\cite{bib15} encode each amino acid as a one-hot vector. This vector is composed of zeros with a single one at the position representing that specific amino acid. One-hot encoding is commonly used in machine learning to transform categorical data into numerical formats that neural networks can process. It is highly effective for deep neural networks to learn the distinctions between amino acid types and capture short- and long-range interactions between amino acids in a sequence. SPOT-1D-Single, the current state-of-the-art single-sequence-based method, achieves a $Q_{3}$ accuracy of 74\%. The superior performance of one-hot encoding over physicochemical encoding underscores that while amino acid side chains play a crucial role in the formation of secondary structures, they are insufficient for fully encoding amino acids.

In this study, we develop two novel chemical representations for twenty types of amino acids by leveraging molecular fingerprints and the dimensionality reduction algorithm FastMap\cite{bib7}. In cheminformatics, a molecular structure can be represented by a graph, where atoms as nodes and chemical bonds between them as edges. Many popular deep neural networks cannot process graph inputs directly and therefore require pre-processing graph inputs into numeric vectors. Morgan fingerprints\cite{mfp} and atom-pair fingerprints\cite{atp} are two commonly used fingerprints that encode graph fragments into fixed-length bit vectors. Although primarily applied to small molecules, these fingerprints have also been adapted to describe peptides\cite{map4}. We employed the FastMap algorithm to reduce the dimensions of these fingerprints, making them computationally feasible for long sequences. The FastMap algorithm is preferred over other dimensionality reduction techniques because it preserves the chemical distances between amino acids. These novel encodings were then tested as inputs for an LSTM-based model. Our results demonstrate that the chemical encodings achieve comparable accuracy to traditional one-hot encoding, confirming their utility as meaningful features for predicting protein secondary structures. Additionally, our chemical encodings are dense vectors of dimensions eighteen and fourteen, compared to 20 for one-hot encoding. This reduction in encoding dimension is especially beneficial for studying long sequences on cost-effective GPUs. Interestingly, a combination of predictions derived from one-hot encoding and our chemical encodings resulted in improved accuracy across all test sets. While physicochemical properties of amino acid side chains can be learned by LSTM models using one-hot encoding\cite{bib12}, the molecular fingerprint-based encodings provide additional, non-redundant information that enhances the model's predictive capabilities. Our method is competitive with the best single-sequence-based models and offers significant computational advantages. It outperforms the LSTM-based model used in SPOT-1D-Single (hereafter referred to as S1D) for most test sets while requiring approximately nine times fewer trainable parameters for each encoding model. These findings demonstrate the importance of chemical details in the formation of protein secondary structures and we hope to encourage more research developments in this direction. 

\section{MATERIALS AND METHODS} 

\subsection{Datasets}
To compare with S1D, we employed the same training, validation, and test sets. The protein sequences, secondary structures, and trained models are available on the authors' website. The training and validation datasets were curated from the ProteinNet dataset\cite{pnet}, which comprises proteins submitted to the Protein Data Bank\cite{pdb} (PDB) before 2016. Originally, these datasets included 39,120 and 100 proteins respectively; however, due to the unavailability of some secondary structure files from the published data, our study utilized 39,119 proteins for training and 98 for validation. Details of the test sets are summarized in Table \ref{tab:testset}.
\vspace{-7pt}
\begin{table}[H]
\captionsetup{justification=justified,singlelinecheck=false} 
\caption{Test sets.} 
\label{tab:testset}
\setlength\tabcolsep{8pt} 
\small
\begin{tabular}{|l | c | c | c | c |} 
\hline
 \textbf{Test Set} & \textbf{Total sequences} & \textbf{Total residues} & \textbf{Max len} & \textbf{Min len}\\ [0.5ex] 
 \hline \hline
 SPOT-2016 & 1466 & 297808 & 1539 & 30\\ 
 SPOT-2016-HQ & 295 & 53592 & 1440 & 30 \\
 SPOT-2018 & 677 & 143318 & 1539 & 30\\
 SPOT-2018-HQ & 125 & 22824 & 1127 & 30\\
 TEST2018 & 250 & 56654 & 615 & 31\\
 NEFF1-2018 & 41 & 4936 & 476 & 31\\
 CASP12-FM & 22 & 7688  & 670 & 106 \\
 CASP13-FM & 17 & 5227 & 552 & 114\\[1ex]
 \hline
\end{tabular}

\bigskip
\raggedright 
Note: The original datasets for SPOT-2016 and SPOT-2018 contained 1,473 and 682 sequences, respectively. However, seven sequences in SPOT-2016 and five sequences in SPOT-2018 were missing secondary structure files. 

Abbreviation: Max len, maximum sequence length; Min len, minimum sequence length. 
\end{table}

SPOT-2016 consists of 1,466 protein sequences released between January 2016 and April 2020. SPOT-2018 is a subset of SPOT-2016, including proteins released after January 2018. SPOT-2016-HQ and SPOT-2018-HQ are high-resolution subsets of their respective datasets, with resolution constraints of less than 2.5 angstroms and R-free values below 0.25. NEFF1-2018 includes 41 sequences from SPOT-2018, all of which have Neff values of 1, indicating proteins with no known homologs. TEST2018, CASP12-FM, and CASP13-FM are three independent test sets. TEST2018 comprises 250 high-resolution sequences that meet the same resolution criteria as SPOT-2016-HQ and SPOT-2018-HQ. CASP12-FM and CASP13-FM consist of 22 and 17 proteins, respectively, released during the CASP12\cite{casp12} and CASP13\cite{casp13} competitions, without known structural templates at the time of release. Together, CASP12-FM, CASP13-FM, and NEFF1-2018 represent three challenging categories for protein structure prediction.

\subsection{Input features}
Each protein sequence is represented by a vector of size $k \times L$, where $L$ is the maximum sequence length of a dataset. For our training data, $L = 4915$. The integer $k$ varies depending on the type of amino acid encoding used: 1) $k = 20$ for one-hot encoding, 2) $k = 18$ for Morgan fingerprint-based encoding, and 3) $k = 14$ for atom-pair fingerprint-based encoding.

Morgan and atom-pair fingerprints encode each amino acid as bit vectors of dimensions 4096 and 8,388,608, respectively. Using the FastMap algorithm, we reduce their dimensions to 18 and 14, respectively. An overview of molecular fingerprints is discussed in Section ~\ref{section:fps} and our implementation of the FastMap algorithm is provided in Section ~\ref{section:fma}.

\subsubsection{Morgan/atompair fingerprints and chemical distance function}
\label{section:fps}
Molecular fingerprints are derived from molecular graphs, representing atoms as nodes and bonds between atoms as edges. These fingerprints are typically represented as bit vectors, where a `1' or `0' indicates the presence or absence of a particular molecular substructure.

The Morgan fingerprint, also known as the circular or extended connectivity fingerprint, is generated by iteratively traversing the molecular graph. This process incorporates information about atom types and bond types within each atom's neighborhood up to a specified radius. 
\vspace{-10pt}
\begin{figure}[h]
\centering
\includegraphics[width=0.35\textwidth]{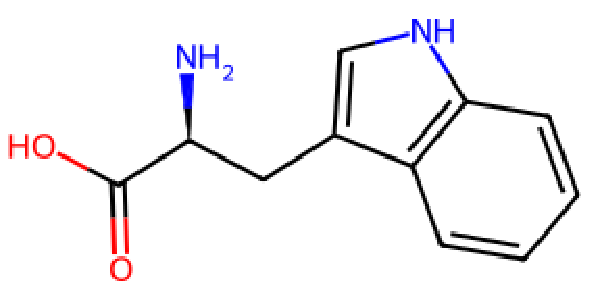}
\caption{Molecular structure of the amino acid Tryptophan.}
\label{fig:tryptophan}
\end{figure}

\begin{figure}[h]
\centering
\includegraphics[width=0.7\textwidth]{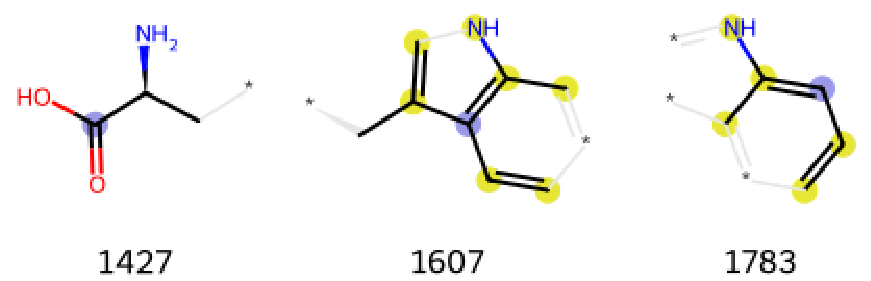}
\vspace{5pt}
\caption{Bit indices 1427, 1607, and 1783 in the Morgan fingerprint representation of Tryptophan correspond to specific substructures. The central atom is highlighted in blue, while aromatic atoms are highlighted in yellow. A search radius of two from the central atom was used.}
\label{fig:bits}
\end{figure}

Figure \ref{fig:tryptophan} represents the molecular structure of the amino acid Tryptophan. Figure \ref{fig:bits} represents the substructures corresponding to bit indices 1427, 1607, and 1783 in the Morgan fingerprint representation of Tryptophan. We set the Morgan fingerprint to be a bit vector of 4096 bits and the radius to be two, which are standard practices in cheminformatics. We used the open-source software RDKit, version 2022.09.4, to generate Morgan fingerprints for all types of amino acids. For Morgan fingerprints, we used the ``GetMorganFingerprintAsBitVect'' function

While Morgan fingerprints excel at capturing local structural features, they may not fully encapsulate global characteristics, such as the overall size and shape of molecules\cite{map4}. In contrast, atom-pair fingerprints are derived by exhaustively enumerating all possible pairs of atoms within a molecule and calculating the shortest paths between them. Unlike Morgan fingerprints, which primarily encode local features, atom-pair fingerprints include more global molecular features by capturing pairwise distances between atoms. For atom-pair fingerprints, we use the ``GetAtomPairFingerprintAsBitVect'' function. This function does not allow the specification of the number of bits; therefore, we used the default number of 8,388,608 bits.

\subsubsection{FastMap alogirthm}
\label{section:fma}
FastMap is an efficient dimensionality reduction algorithm that transforms complex objects into points within a low-dimensional Euclidean space while maintaining the original distances between each pair of objects. This method is particularly useful when feature extraction is challenging, but distance functions are easily obtained. FastMap has found applications in various domains, such as text mining, bioinformatics, and data visualization. 

\begin{figure}[H]
\centering
\begin{subfigure}{0.48\textwidth}
\centering
\includegraphics[width = \textwidth]{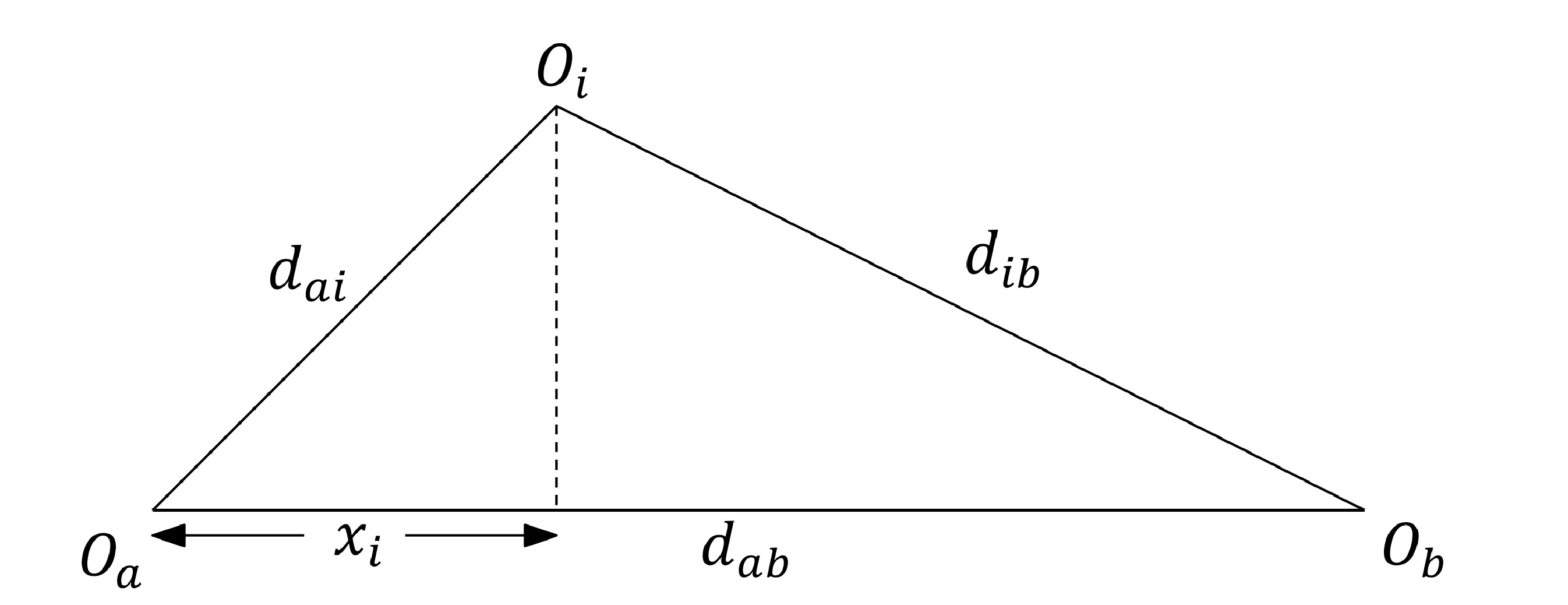}
\caption{Step 2: Cosine law}
\label{fig:left}
\end{subfigure}
\begin{subfigure}{0.48\textwidth}
\centering
\includegraphics[width = \textwidth]{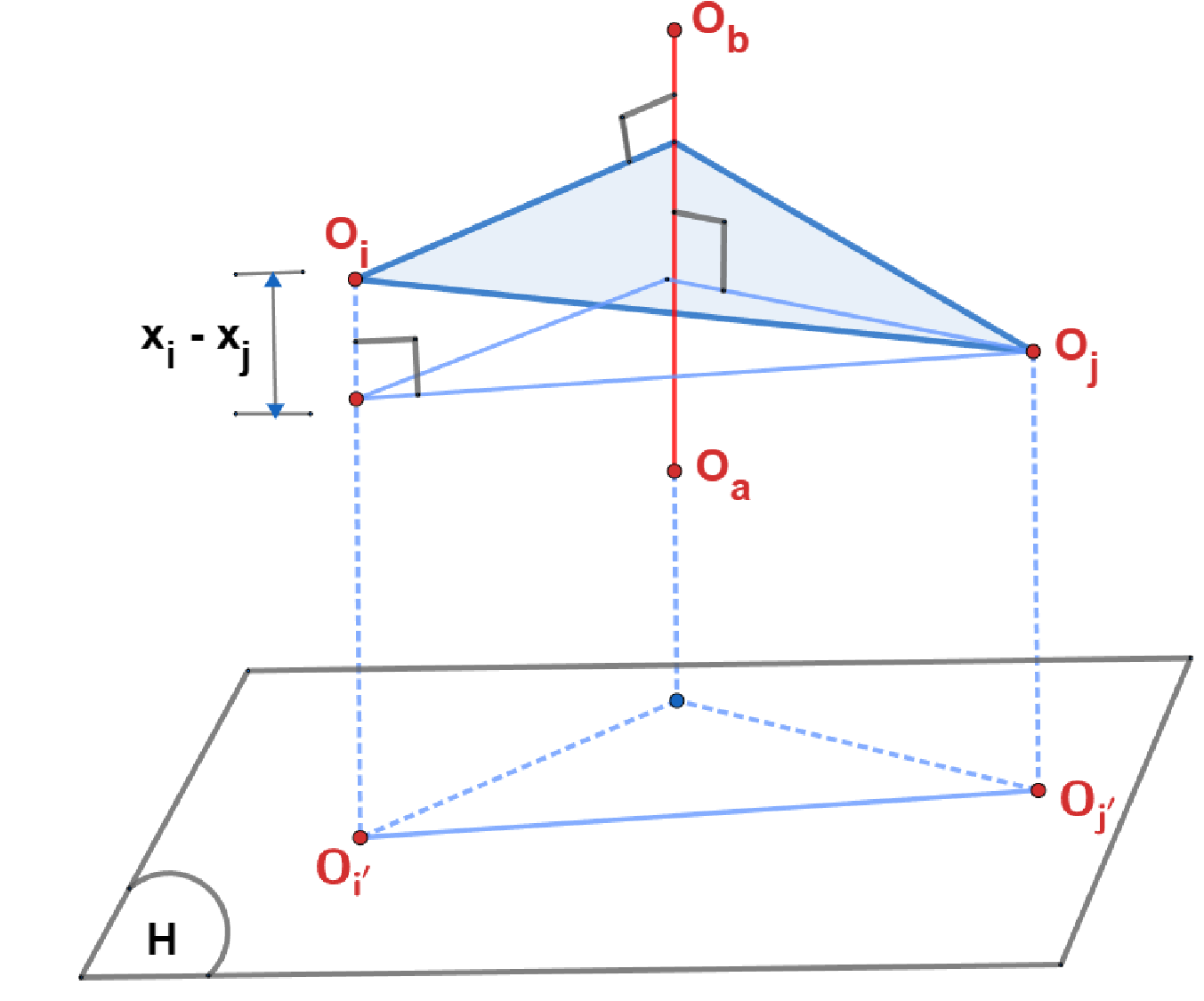}
\caption{Step 3: Update distance}
\label{fig:right}
\end{subfigure}
\caption{FastMap algorithm. Figure (A): Step 2 of Algorithm~\ref{box:example}. Figure (B): Step 3 of Algorithm~\ref{box:example}.}
\label{fig:fm}
\end{figure}

The pseudocode for our implementation of the FastMap algorithm to reduce the dimension of molecular fingerprints is provided in Algorithm~\ref{box:example}.
\begin{boxA}[label={box:example}]
\textbf{Input}: 
\vspace{-4pt}
\begin{itemize}
\item[] 
    \begin{itemize}
    \item[$-$] 20 amino acids fingerprints $\{O_{1}, O_{2}, \ldots, O_{20}\}$
    \item[$-$] Dice distance function $\mathcal{D}$
    \item[$-$] Stopping conditions: embedding dimension $K = 100$, threshold $\epsilon = 0.0001$    
    \end{itemize}
\end{itemize}

\textbf{For $\mathbf{k = 1, 2, \ldots, K}$}:
\vspace{-4pt}
\begin{itemize}
  \item[] Step 1: Heuristically find a pair of objects, $O_{a}, O_{b}$, such that $\mathcal{D}(O_{a}, O_{b})$ is maximized.
\end{itemize}
\hspace{0.95cm}\textbf{While $\mathbf{\mathcal{D}(O_{a}, O_{b})}$ > $\mathbf{\epsilon}$, do: }
\vspace{-4pt}
  \begin{itemize}
    \item[]
        \begin{itemize}              
              \item[] Step 2: Calculate the $k^{th}$ coordinates for every object using the Cosine Law:
                  \begin{equation*}
                      x_{i} = \frac{\mathcal{D}(O_{a}, O_{i})^{2} + \mathcal{D}(O_{a}, O_{b})^{2} - \mathcal{D}(O_{i}, O_{b})^{2}}{2\mathcal{D}(O_{a}, O_{b})}
                  \end{equation*}
                  where $i = 1,2, \ldots, 20$.
              \item[] Step 3: Update the distance function for all pairs of objects:
              \begin{equation*}
                  \mathcal{D}^{\prime}(O_{i^{\prime}}, O_{j^{\prime}})^{2} = \mathcal{D}(O_{i}, O_{j})^{2} - (x_{i} - x_{j})^{2}
              \end{equation*}
                
        \end{itemize}
\end{itemize}
\end{boxA}
The inputs to the FastMap algorithm are 20 fingerprints for 20 types of amino acids and the Dice distance function\cite{Dice1945}. We represent Morgan/atom-pair fingerprints as points $O_{i}$ in a high-dimension space (dimensions of 4096 for Morgan fingerprints and 8,388,608 for atom-pair fingerprints). The algorithm begins by selecting initial pivot points that are farthest apart and computing their distances to all other points. It then projects these points onto a line defined by the chosen pivot points and calculates the projection using the Cosine Law. Figure \ref{fig:fm}(A) represents the projection of an arbitrary $O_{i}$ point onto the line defined by two pivot points $O_{a}$ and $O_{b}$. The projection $x_{i}$ is calculated by the Cosine Law, where $d_{ai}$, $d_{ab}$, $d_{ib}$ are the Dice distances. This projection, illustrated in Figure \ref{fig:fm}(A), is used to update the distances between points based on the Pythagorean theorem, as shown in Figure \ref{fig:fm}(B). Using these updated distances, the algorithm selects a new pair of pivot points for the next iteration. This iterative process continues, recalculating distances and redefining pivot points, until the embedding reaches the desired dimensionality ($K=100$) or the distance between the farthest pair of points drops below a threshold ($\epsilon = 0.0001$)\cite{e}.
\vspace{5pt}

\vspace{-5pt}
\vspace{-10pt}
Note that the distance function $\mathcal{D}$ in the FastMap algorithm does not need to follow the triangle inequality. It only needs to satisfy 3 conditions: 1) $\mathcal{D}(O_{1}, O_{2}) = \mathcal{D}(O_{2}, O_{1})$; 2) $\mathcal{D} (O_{1}, O_{2}) >= 0$; 3) $\mathcal{D} (O_{1}, O_{1}) = 0$. Examples of distance functions are the edit distance for DNA sequences\cite{Levenshtein1966, Gusfield1997} and the cross-correlation for earthquake signals\cite{fmsvm}. In this study, we use the Dice distance. Given two binary vectors $\mathbf{A}$ and $\mathbf{B}$, the distance can be defined as: 
\vspace{-3pt}
\begin{equation}
    \mathcal{D}(\mathbf{A}, \mathbf{B}) = 1 - \frac{2 \times |\mathbf{A} \cap \mathbf{B}|}{|\mathbf{A}| + |\mathbf{B}|}
\end{equation}
, where \(|\mathbf{A}|\) and \(|\mathbf{B}|\) represents the cardinality of set \(\mathbf{A}\) and set \(|\mathbf{B}|\), respectively, and \(|\mathbf{A} \cap \mathbf{B}|\) represents the number of elements common to both sets \(\mathbf{A}\) and \(\mathbf{B}\).

\subsection{Outputs}
The outputs are the three-state secondary structures and eight-state secondary structures. The DSSP program (Define Secondary Structure of Proteins)\cite{bib16} defines eight secondary structures: $3_{10}$-helix (G), $\alpha$-helix (H), $\pi$-helix (I), $\beta$-bridge (B), $\beta$-strand (E), high curvature loop (S), $\beta$-turn (T), and coil (C). We then simplified these eight states into three states: sheet (E), helix (H), and coil (C), using the same conversion method as SPOT-1D-Single. Specifically, B and E are merged into sheet (E); G, H, and I are merged into helix (H); and all other states are classified as coil (C).

\subsection{Evaluation metrics}
The most commonly used metrics for protein secondary structure prediction are the 3-state, $Q_{3}$, and the 8-state accuracy, $Q_{8}$. These metrics calculate the percentage of correctly predicted secondary structures as follows:
\[
Q_{k} = 100\% \times \frac{\sum_{s\in S}^{}n_{s}}{N} 
\]
where $k = \{3, 8\}$, $N$ is the total number of residues, and $n_{s}$ is the number of correctly predicted residues in state $s$.  For 3-state predictions, $S=\{H, E, C\}$; for 8-state predictions, $S=\{C, E, T, S, H, G, B, I\}$.

$Q_{3}$ and $Q_{8}$ accuracies can be calculated both at the residue level and the sequence level. At the residue level, $N$ and $n_{s}$ include all amino acids across different protein sequences. At the sequence level, these metrics are computed for each protein sequence individually and then averaged across all sequences. 

While the prediction is made for all amino acids, only amino acids with defined secondary structures are counted to $Q_{3}$ and $Q_{8}$ accuracies.

\subsection{Multi-task learning for 3-state and 8-state predictions}
\begin{figure}[h]
\centering
\includegraphics[width=0.65\textwidth]{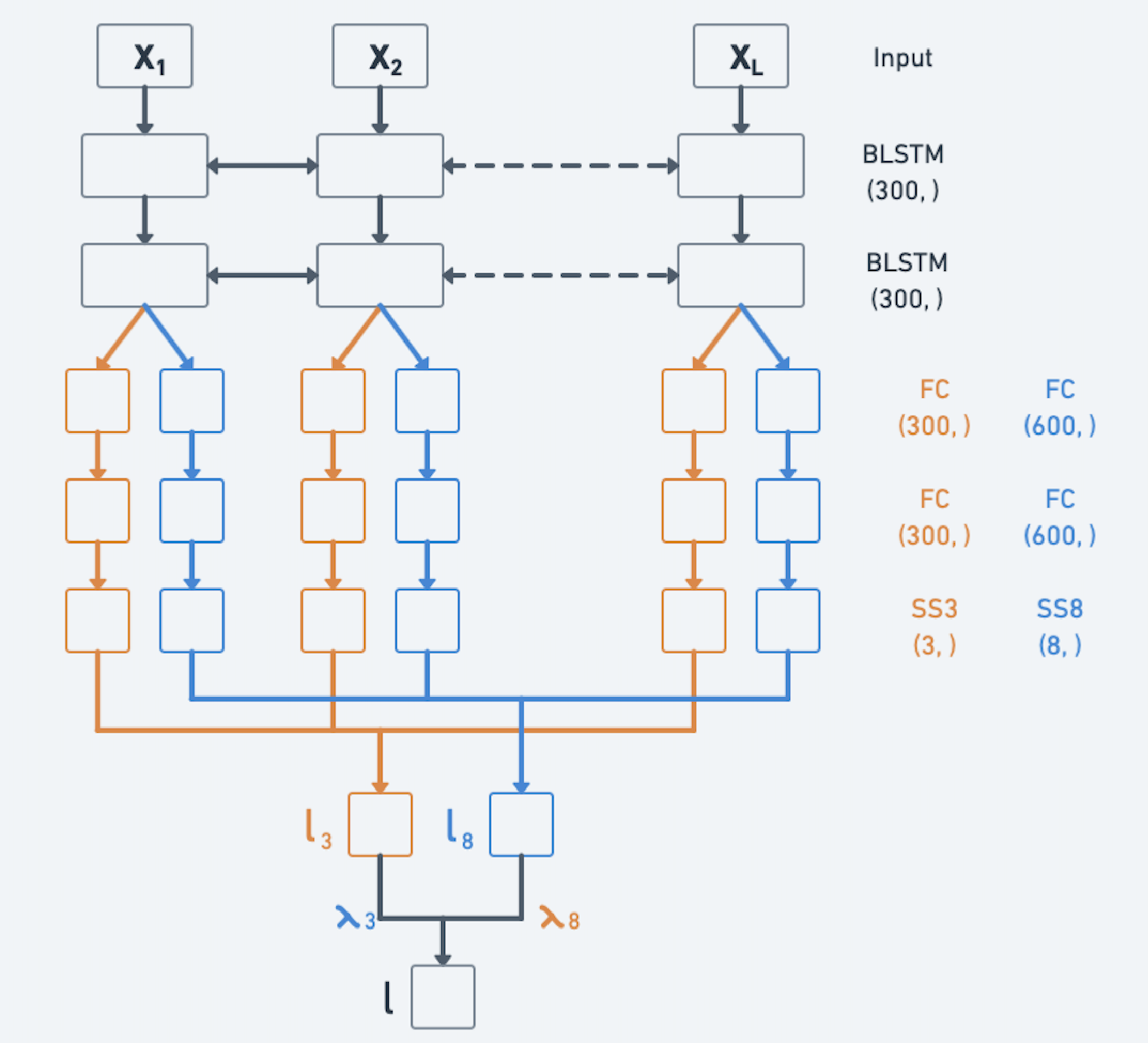}
\caption{Long short-term memory-based architecture for three-state and eight-state prediction. $l_{3}$ and $l_{8}$ are the cross-entropy losses for 3-state and 8-state prediction tasks, $\lambda_{3}$ and $\lambda_{8}$ are their corresponding weights, $l$ is the total loss. BLSTM denotes the bidirectional long short-term memory layer and FC denotes the fully connected layer. SS3 and SS8 denote the outputs for 3-state and 8-state structures, respectively.}
\label{fig:model}
\end{figure}

The LSTM-based model described in the SPOT-1D-Single paper includes two bidirectional LSTM layers, each with hidden dimensions of 1,000 per direction, followed by two fully connected layers, each of size 2,000. Our multi-task learning architecture, illustrated in Figure \ref{fig:model}, is a refined version of the SPOT-1D-Single model. It comprises two bidirectional LSTM layers, each with hidden dimensions of $n$ per direction. Like SPOT-1D-Single, we incorporate a dropout layer with a rate of 0.5 after each LSTM layer to prevent overfitting. In contrast to earlier designs, where two tasks share the LSTM and fully connected layers, our model provides separate fully connected layers for each prediction task. This setup allocates more neurons to the 8-state prediction task due to its higher complexity. Specifically, the fully connected layers for the 3-state and 8-state prediction tasks have sizes $n$ and $2n$, respectively. Our ensemble approach uses $n = 300$ for all encodings. This value is found by a grid search procedure discussed in Session~\ref{sec:ensemble}. The output architecture includes three nodes for the 3-state prediction task and eight for the 8-state prediction task, totaling 11 output nodes in our model.

The training was conducted using a batch size of 8 and a learning rate of 0.001, limiting the training to a maximum of 15 epochs. The models were saved after each training epoch. Training was prematurely halted if $Q_3$ accuracies on the validation data did not improve over five consecutive epochs. Sequences shorter than 4915 residues were zero-padded. A custom cross-entropy loss function was implemented, excluding amino acids with undefined secondary structures and padded residues from the loss calculation.

The overall loss is calculated as a weighted sum of the losses for the two tasks:
\begin{equation*}
    l = \lambda_{3}l_{3} + \lambda_{8}l_{8}
\end{equation*}
where $l_{3}$ and $l_{8}$ represent the cross-entropy losses for the 3-state and 8-state prediction tasks, respectively. In our study, the weights $\lambda_{3}$ and $\lambda_{8}$ were both set to 0.5.

We implemented our model using TensorFlow 2.12. Training was accelerated by CUDA 11.8 and cuDNN 8.6. 

\subsection{Ensemble learning of one-hot and chemical encodings}
\label{sec:ensemble}
Our objective is to demonstrate that an ensemble model combining models trained using Morgan fingerprint-based encoding (MFE model), atom-pair fingerprint-based encoding (AFE model), and one-hot encoding (OHE model) can provide additional information that an LSTM-based model might not capture using one-hot encoding alone. To isolate the effect of these different encoding methods, we employ the same LSTM neural network architecture for all three encodings. This approach ensures that any performance improvements in the ensemble over the OHE model can be attributed directly to the encoding methods rather than variations in the network structure.

To identify the optimal number of $n$ for the ensemble, we performed a grid search with $n$ values set at 200, 250, 300, 350, and 400. We limited our search to a maximum of 400 due to computational resource constraints. For each value of $n$, we identified the best-performing model during training for each encoding method based on the highest $Q_3$ accuracy observed in the validation dataset. For each value of $n$, we aggregated the predictions from the best models by averaging the probabilities assigned to each class of secondary structures:
\vspace{-8pt}
\begin{equation*}
P_{S} = \frac{P_{S}^{\text{OHE}} + P_{S}^{\text{MFE}} + P_{S}^{\text{AFE}}}{3}
\end{equation*}
, where $S=\{H, E, C\}$ for 3-state predictions and $S=\{C, E, T, S, H, G, B, I\}$ for 8-state predictions. The secondary structure type with the highest prediction probability is then predicted as $S^{*} = \underset{S}{\text{argmax}} \, \{P_{S}\}$.

Table \ref{tab:tuning} displays the $Q_{3}$ accuracies for each encoding model and for the ensemble model across five values of LSTM hidden dimensions. 
\begin{table}[h]

\centering
\caption{$Q_{3}$ accuracy (\%) on the validation dataset for OHE, MFE, AFE and ensemble models for different values of $n$}
\label{tab:tuning}
\setlength\tabcolsep{8pt}
\begin{tabular}{|l | c | c | c | c |} 
\hline
\textbf{n} & \textbf{OHE} & \textbf{MFE} & \textbf{AFE} & \textbf{Ensemble}\\
\hline 
200 & 72.53 & 72.38 & 72.20 & 72.86\\ 
250 & 72.58 & 72.05 & 72.03 & 72.92\\
300 & 72.44 & 72.54 & 72.36 & 73.22\\
350 & 72.33 & 72.25 & 72.63 & 73.19\\
400 & 72.54 & 72.35 & 72.22 & 73.13\\
\hline
\end{tabular}
\end{table}

MFE and AFE models perform comparably to OHE model across all values of $n$. Notably, the ensemble consistently improves the accuracies for individual models for all values of $n$. With $n = 300$, the ensemble achieves the highest $Q_3$ accuracy on the validation dataset. In the following sessions, we will report the results obtained using the LSTM-based model with hidden dimension $n = 300$.

\vspace{10pt}
\section{RESULTS}

\subsection{Amino acids in the embedding space}
\begin{figure}[h]
\centering
\includegraphics[width=\textwidth]{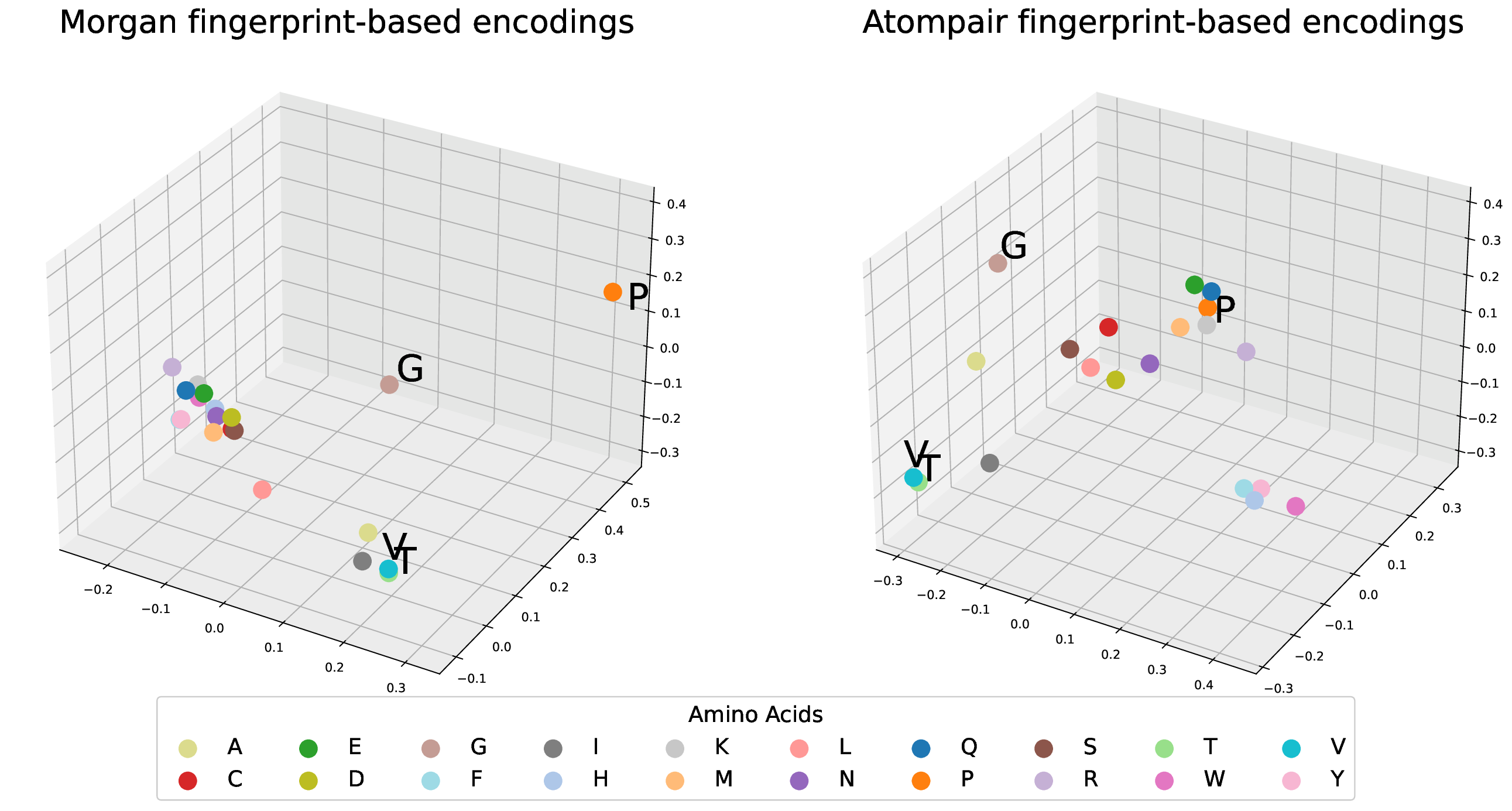}
\caption{Spatial distribution of amino acids in Euclidean space for Morgan fingerprint-based encoding and atom-pair fingerprint-based encoding.}
\label{fig:aa_space}
\end{figure}

FastMap maps the Morgan fingerprints and the atom-pair fingerprints of amino acids to points in Euclidean spaces of dimensions 18 and 14, respectively. This approach enables visualization of the spatial distribution of the 20 amino acids within the embedding spaces. Figure \ref{fig:aa_space} illustrates these distributions for Morgan fingerprint-based and atom-pair fingerprint-based encodings, with each amino acid's coordinates represented by the three principal components derived from PCA analysis.

\begin{figure}[h]
\centering
\includegraphics[width=\textwidth]{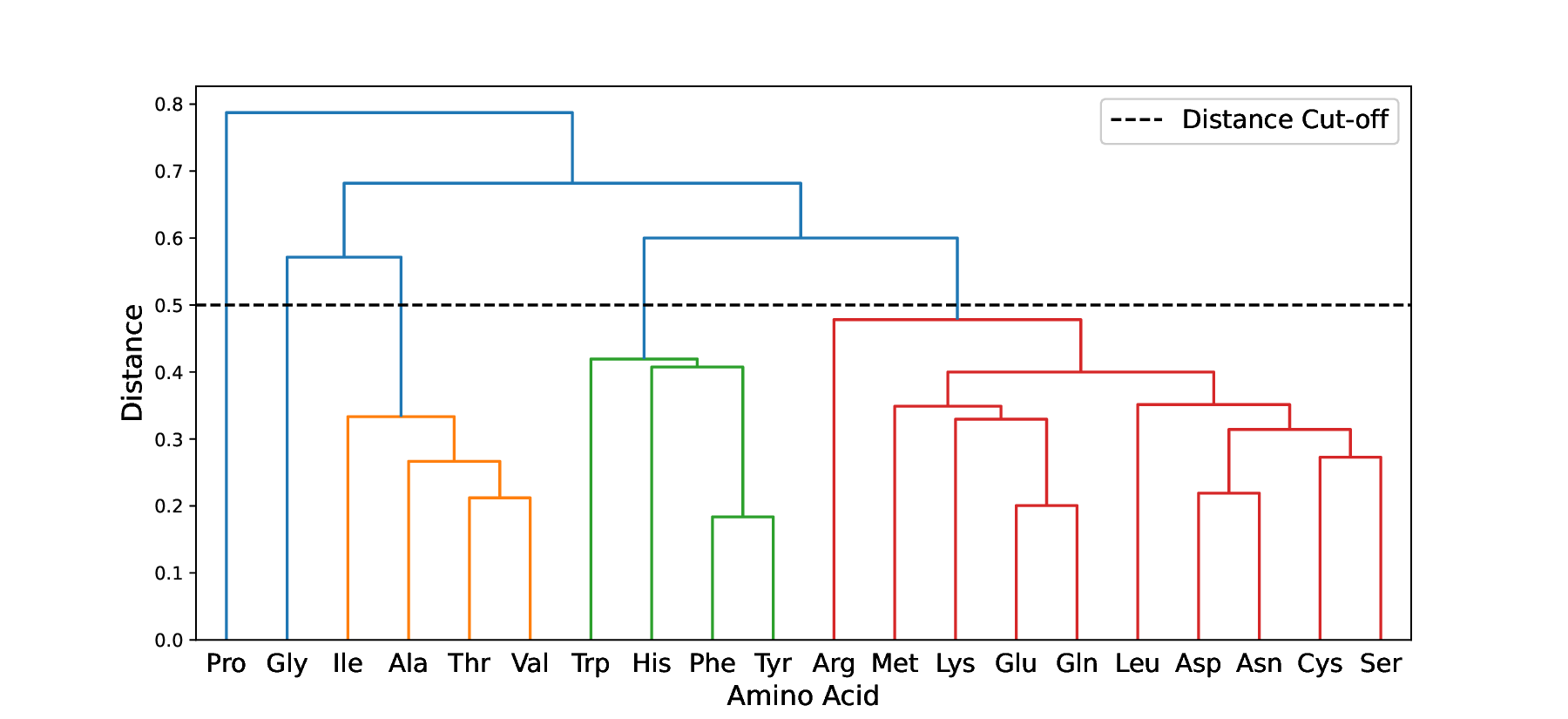}
\caption{Hierarchical Clustering Dendrogram for Morgan fingerprint-based encoding.}
\label{fig:dendogram_morgan}
\end{figure}

\begin{figure}[h]
\centering
\includegraphics[width=\textwidth]{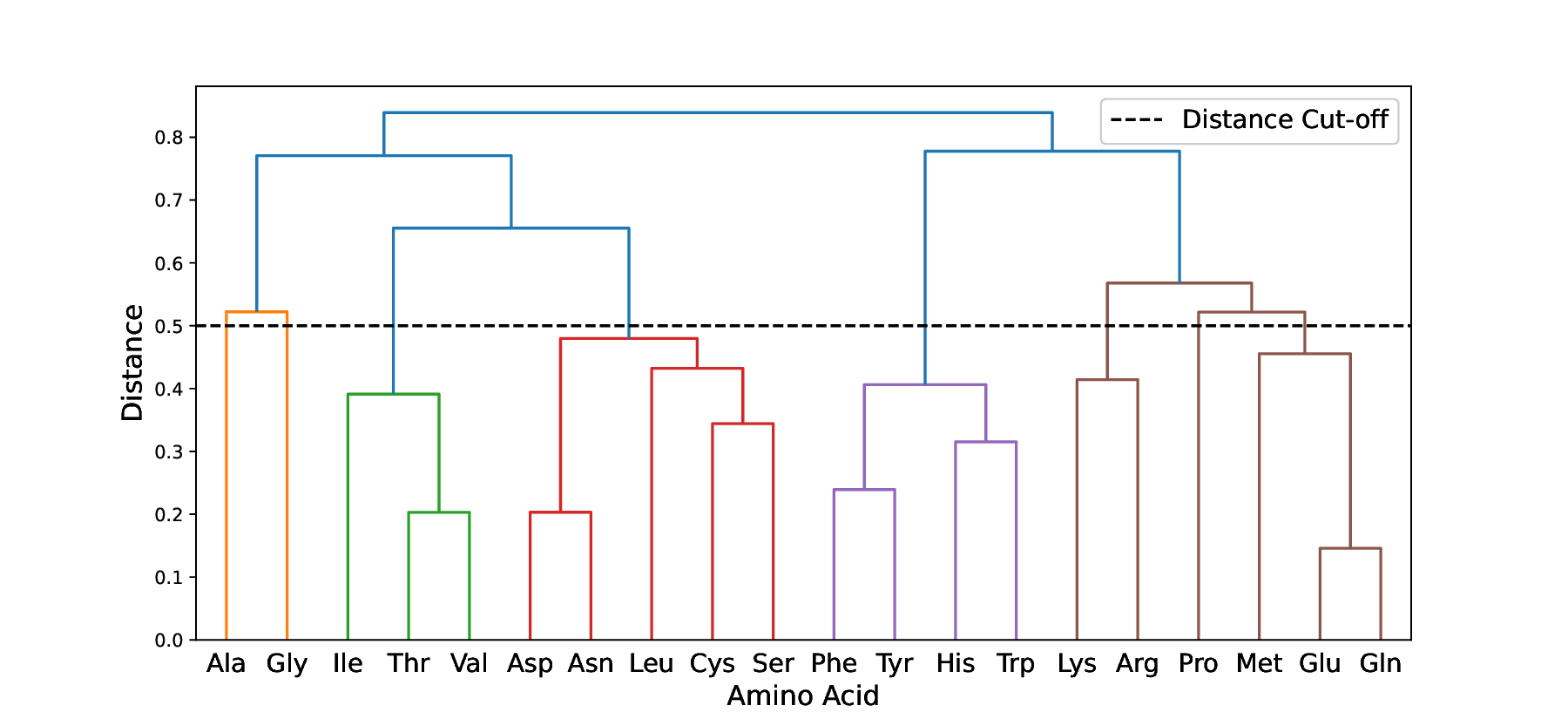}
\caption{Hierarchical Clustering Dendrogram for atom-pair fingerprint-based encoding.}
\label{fig:dendogram_atp}
\end{figure}

 Figure \ref{fig:dendogram_morgan} and figure \ref{fig:dendogram_atp} show the results of complete-linkage clustering for 20 types of amino acids using Morgan fingerprint-based and atom-pair fingerprint-based encodings. We use Euclidean distance and a distance cut-off of 0.5 to stop the clustering process. Clusters consist of amino acids that are proximate in the embedding Euclidean space. Since FastMap preserves the chemical distances between amino acids, proximity in the embedding space reflects similarities in their molecular structures.

For Morgan fingerprint-based encodings, there are five clusters: (Gly), (Pro), (Phe, His, Trp, Tyr), (Ala, Ile, Thr, Val), and (Cys, Glu, Asp, Lys, Met, Leu, Asn, Gln, Ser, Arg). For atom-pair fingerprint-based encodings, there are eight clusters: (Gly), (Pro), (Phe, His, Trp, Tyr), (Ala), (Ile, Thr, Val), (Cys, Asp, Leu, Asn, Ser), (Lys, Arg), (Glu, Met, Gln). Morgan and atom-pair fingerprints use different methods to encode the structural information of amino acids, which can result in different molecular substructures being captured. Notably, both encoding methods agree on three specific clusters: (Gly), (Pro), and (Phe, His, Trp, Tyr). This consistency can be attributed to the structural uniqueness of these amino acids, making them be distinguished easily from the rest by both fingerprint methods. The remaining amino acids share many structural similarities and are more challenging to differentiate. As a result, different clustering patterns are formed for Morgan and atom-pair fingerprint-based encodings.
\vspace{-10pt}

A careful examination of all molecular substructures for each amino acid is needed to explain the formation of each cluster. However, for some clusters, the structural similarity between their members can be inspected easily visually. For example, Gly is isolated in its cluster due to its simple structure as the only amino acid without a side chain. Pro is distinct for its unique cyclic structure, where a secondary amine is bonded directly to the alpha carbon. Phe, His, Trp, and Tyr are the only four standard amino acids with aromatic side chains. Lys and Arg are the two longest linear amino acids with extended side chains. The cluster of Lys and Arg indicates the effectiveness of atom-pair fingerprinting in capturing the overall shape and size of molecules.

\subsection{Ensemble performance}
\subsubsection{Accuracy gain over OHE model}
\label{sec:ensemble_performance}
Table \ref{tab: individual_reps} presents $Q_{3}$ and $Q_{8}$ accuracies obtained using MFE, OHE, and AFE models across eight test sets. The metrics are calculated at the residue level. We also display the ensemble's performance obtained using the procedure described in Session~\ref{sec:ensemble} and its accuracy gains over OHE model.
\begin{table}[h]
\centering
\caption{Performance of individual models and accuracy improvement of the ensemble over OHE model across eight test sets.}
\label{tab: individual_reps}

\begin{tabular}{llrc@{\quad}rc@{\quad}rc@{\quad}rc@{\quad}}
\toprule
 &  & \multicolumn{2}{c}{SPOT-2016} &
\multicolumn{2}{c}{SPOT-2016-HQ} &
\multicolumn{2}{c}{SPOT-2018} & \multicolumn{2}{c}{SPOT-2018-HQ} \\
\cmidrule(lr){3-4}\cmidrule(lr){5-6}\cmidrule(lr){7-8}\cmidrule(lr){9-10}
 &  & Q3 & Q8 & Q3 & Q8 & Q3 & Q8 & Q3 & Q8 \\
\midrule
OHE  & & 73.37 & 60.2 & 71.96 & 59.79 & 72.76 & 58.9 & 70.29 & 57.75 \\
AFE  & & 73.39 & 59.86 & 72.78 & 60.38 & 72.72 & 58.54 & 70.94 & 58.15\\
MFE  & & 73.83 & 60.59 & 72.73 & 60.41 & 73.24 & 59.22 & 70.71 & 57.78 \\
Ensemble & & 74.37 & 61.16 & 73.42 & 61.29 & 73.77 & 59.85 & 71.62 & 59.12 \\
Gain & & +1.00 & +0.96 & +1.46 & +1.50 & +1.01 & +0.95 & +1.33 & +1.37 \\
\bottomrule
\end{tabular}

\vspace{1em} 

\begin{tabular}{llrc@{\quad}rc@{\quad}rc@{\quad}rc@{\quad}}
\toprule
 &  & \multicolumn{2}{c}{TEST2018} &
\multicolumn{2}{c}{NEFF1-2018} &
\multicolumn{2}{c}{CASP12-FM} & \multicolumn{2}{c}{CASP13-FM} \\
\cmidrule(lr){3-4}\cmidrule(lr){5-6}\cmidrule(lr){7-8}\cmidrule(lr){9-10}
 &  & Q3 & Q8 & Q3 & Q8 & Q3 & Q8 & Q3 & Q8 \\
\midrule
OHE   & & 72.99 & 60.91 & 74.24 & 63.36 & 71.36 & 57.77 & 70.71 & 57.67 \\
AFE  & & 72.95 & 60.67 & 74.98 & 62.85 & 71.08 & 56.84 & 71.34 & 58.27 \\
MFE  & & 73.18 & 60.83 & 75.15 & 64.08 & 71.60  & 58.41 & 71.74  & 59.18 \\
Ensemble & & 74.03 & 61.91 & 75.54 & 64.12 & 72.38 & 58.56 & 71.82 & 60.06 \\
Gain & & +1.04 & +1.00 & +1.30 & +0.76 & +1.02 & +0.79 & +1.11 & +2.39 \\
\bottomrule
\end{tabular}
\end{table}

The performances of the three different encodings are on a similar level. Although there are slight variations between them, none of the encoding methods consistently outperforms the others across all datasets by a large margin. Hence, we cannot conclude that one encoding method is markedly better than the others. The ensemble method consistently outperforms the individual models across all test sets, proving that the chemical-based representations, Morgan fingerprint-based encoding and atom-pair fingerprint-based encoding, can provide additional information that an LSTM-based model could not learn from one-hot encoding alone. The gains are typically around 1\% for $Q_{3}$ and varying more broadly for $Q_{8}$. The most substantial gains were observed in the challenging dataset CASP13-FM, where the \(Q_8\) accuracy increased significantly by 2.39\%, and in the SPOT-2016-HQ dataset, which saw an improvement of 1.46\%  in \(Q_3\) accuracy. 

\subsubsection{Accuracy improvement for amino acid types}
\begin{figure}[h]
\centering
\includegraphics[width=0.7\textwidth]{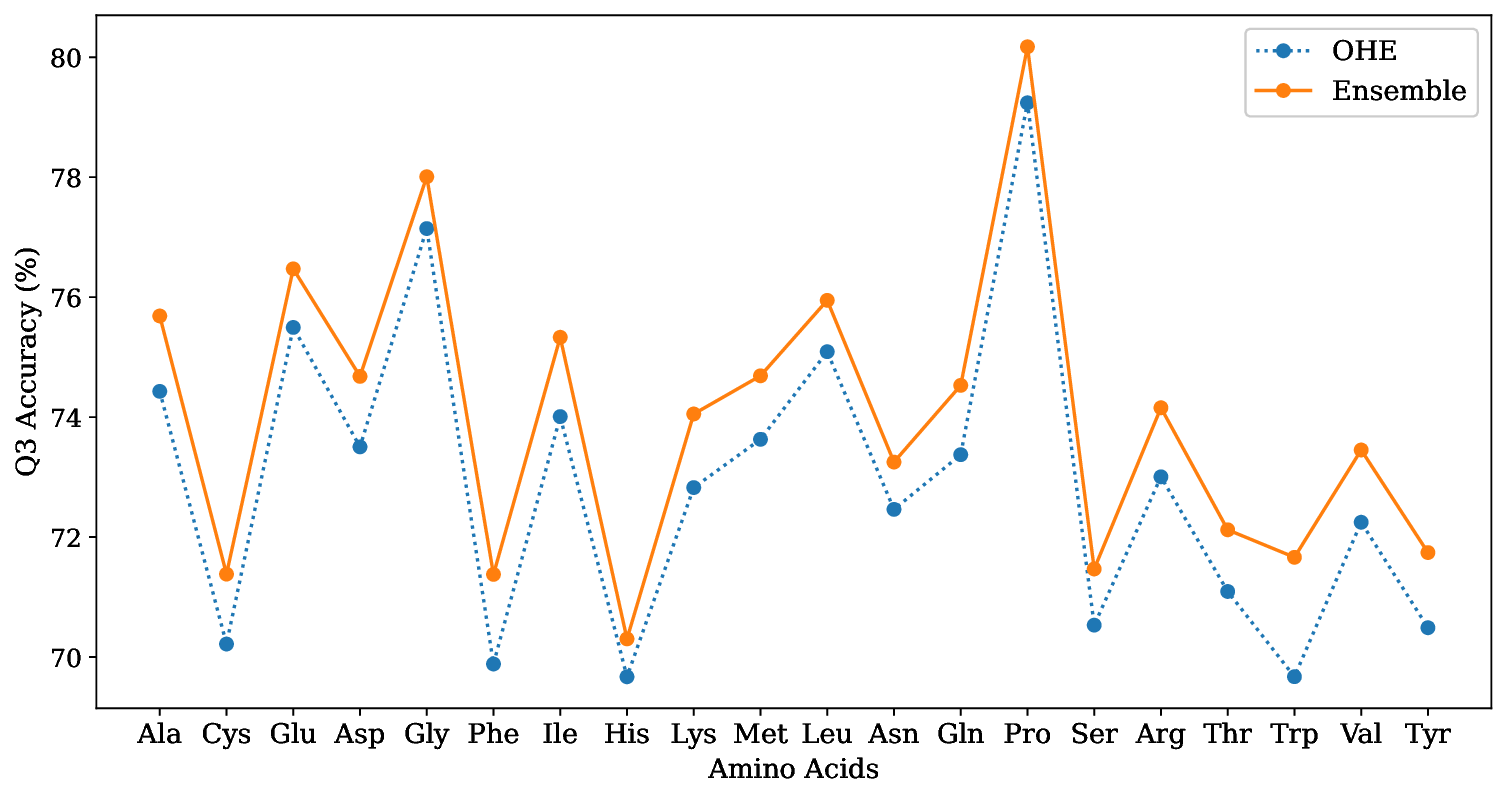}
\caption{$Q_{3}$ accuracy for individual amino acids for the ensemble model and the OHE model on the SPOT-2016 dataset.}
\label{fig:aas_ensemble_ohe}
\end{figure}

Figure \ref{fig:aas_ensemble_ohe} represents $Q_{3}$ accuracy for each type of amino acid on the largest test set SPOT-2016. We represent the results for the ensemble and the OHE model. The ensemble model consistently improves the accuracy for all amino acid types, suggesting the effectiveness of our ensemble model in capturing the interactions between amino acids in protein sequences. The most significant accuracy improvement is seen for the rare amino acid Tryptophan, where the ensemble achieves about $2\%$ better in $Q_{3}$ accuracy. 

\begin{figure}[h]
\centering
\includegraphics[width=0.7\textwidth]{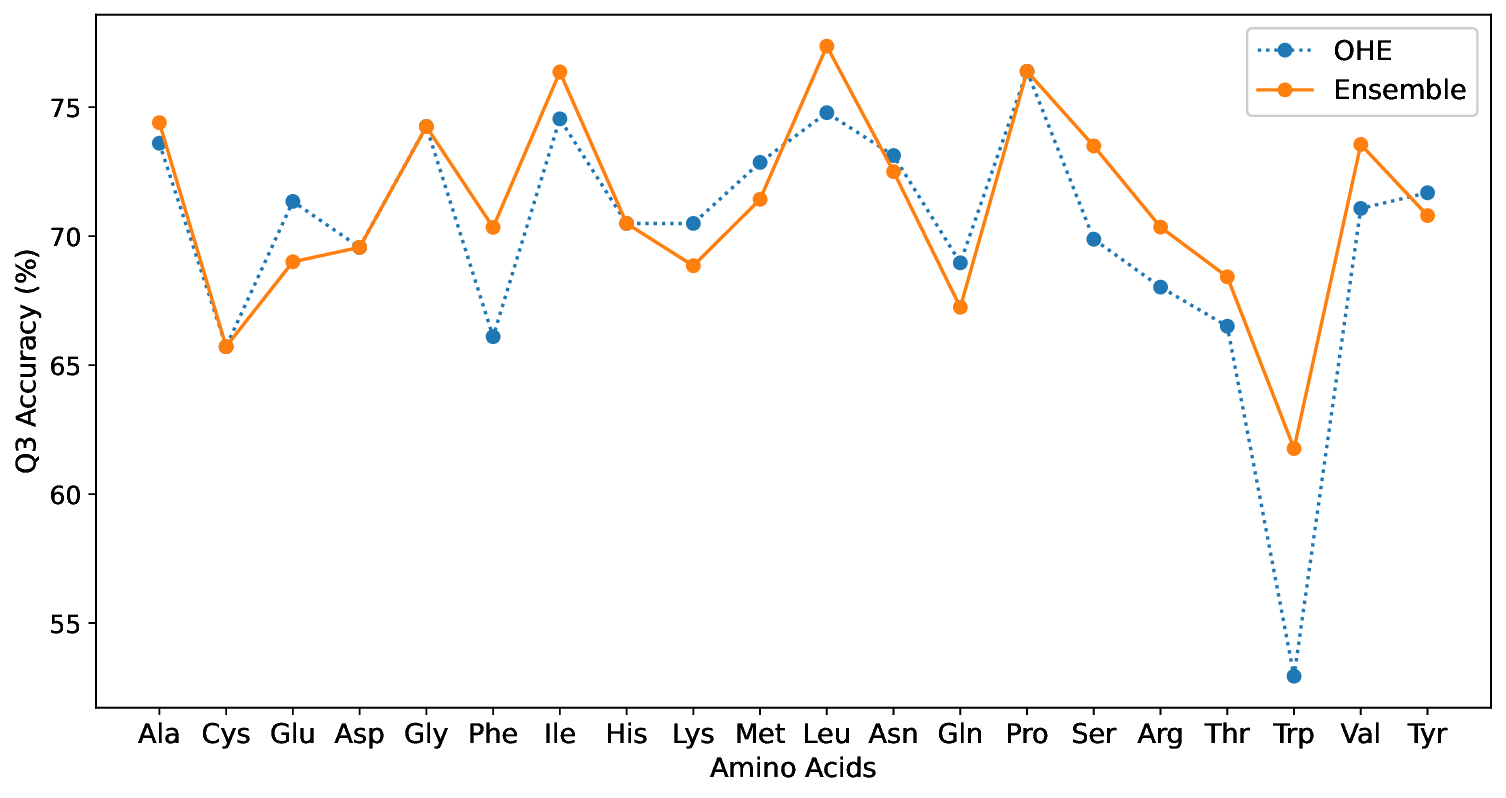}
\caption{$Q_{3}$ accuracy for individual amino acids for the ensemble model and the OHE model on the CASP12-FM dataset.}
\label{fig:aas_ensemble_ohe_casp12}
\end{figure}

For the TEST2018 dataset, the ensemble model also improves the $Q_{3}$ accuracy for all amino acid types (Figure S2). For the challenging datasets CASP12-FM, CASP13-FM, and NEFF1-2018, the accuracy improvement is also seen for most amino acid types. Figure \ref{fig:aas_ensemble_ohe_casp12} shows the result for the CASP12-FM dataset. Similar to SPOT-2016, CASP12-FM also sees the most significant improvement for Tryptophan, a remarkable increase of 8.82\% in $Q_{3}$ accuracy. The results for CASP13-FM and NEFF1-2018 are provided in the supplementary (Figure S4, S5). 

\newpage
\subsubsection{Accuracy for coil, sheet and helix structures}
\begin{table}[h]
\centering
\caption{F1 scores for coil, sheet, and helix using the OHE and the ensemble model.}
\label{tab:F1}

\setlength{\tabcolsep}{3pt} 

\begin{tabular}{l@{\hspace{10pt}}lcccccc} 
\toprule
3-state & Model & SPOT-2016 & TEST2018 & CASP12-FM & CASP13-FM & NEFF1-2018 \\

\midrule
Coil & OHE & 71.58 & 71.77 & 70.28 & 70.37 & 71.74\\
  & Ensemble & 72.84 & 72.85 & 71.76 & 71.81 & 74.00\\
Sheet & OHE & 57.99 & 66.10 & 61.65 & 60.93 & 63.09\\
  & Ensemble & 59.39 & 67.42 & 63.12 & 61.55 & 61.39\\
Helix & OHE & 79.43 & 78.32 & 78.48 & 76.17 & 79.98\\
  & Ensemble & 80.18 & 79.30 & 79.08 & 77.12 & 81.38\\
         
\bottomrule
\end{tabular}
\end{table}

Table \ref{tab:F1} displays F1 scores for coil, sheet, and helix 
structures, demonstrating that the ensemble model outperforms the OHE model across various datasets. Specifically, in SPOT-2016, TEST2018, CASP12-FM, and CASP13-FM, the ensemble model achieves higher accuracy for all three structure types. However, in the NEFF1-2018 dataset, while the ensemble model performs better in classifying coil and helix structures, it underperforms relative to the OHE model in classifying sheet structures. The performance enhancements of the ensemble model over the OHE model are moderate for SPOT-2016 and TEST2018. Notably, the most significant improvement for coil structures—a 2.26\% increase—occurs in NEFF1-2018. The most notable improvements for sheet structures are observed in CASP12-FM, while the most significant gains for helix structures are found in NEFF1-2018. These findings highlight the benefits of the ensemble model, especially in classifying specific secondary structures found in challenging datasets like proteins with limited homologs or those with no known structural templates.

\subsubsection{Comparison with the latest LSTM-based model}
In the original SPOT-1D-Single study, the authors reported the accuracy of their LSTM-based model (S1D) for the SPOT-2016, TEST2018, CASP12-FM, and CASP13-FM datasets. We accessed the trained model through the authors' website and successfully reproduced the results for the TEST2018, CASP12-FM, and CASP13-FM datasets. However, our results for the SPOT-2016 dataset are slightly different from those reported in the original study, attributable to the exclusion of seven sequences from our analysis due to the unavailability of corresponding DSSP files. 
\begin{table}[h]
\centering
\caption{Comparative performance of the ensemble model and the S1D model across eight test sets.}
\label{tab: s1d_comp}

\begin{tabular}{llrc@{\quad}rc@{\quad}rc@{\quad}rc@{\quad}}
\toprule
 &  & \multicolumn{2}{c}{SPOT-2016} &
\multicolumn{2}{c}{SPOT-2016-HQ} &
\multicolumn{2}{c}{SPOT-2018} & \multicolumn{2}{c}{SPOT-2018-HQ} \\
\cmidrule(lr){3-4}\cmidrule(lr){5-6}\cmidrule(lr){7-8}\cmidrule(lr){9-10}
 &  & Q3 & Q8 & Q3 & Q8 & Q3 & Q8 & Q3 & Q8 \\
\midrule
S1D  & & 73.37 & 60.57 & 73.07 & 61.05 & 72.73 & 59.08 & 71.53 & 58.99 \\
Ensemble   & & 74.37 & 61.16 & 73.42 & 61.29 & 73.77 & 59.85 & 71.62 & 59.12 \\
Difference & & +1.00 & +0.59 & +0.35 & +0.24 & +1.04 & +0.77 & +0.09 & +0.13 \\

\bottomrule
\end{tabular}

\vspace{1em} 

\begin{tabular}{llrc@{\quad}rc@{\quad}rc@{\quad}rc@{\quad}}
\toprule
 &  & \multicolumn{2}{c}{TEST2018} &
\multicolumn{2}{c}{NEFF1-2018} &
\multicolumn{2}{c}{CASP12-FM} & \multicolumn{2}{c}{CASP13-FM} \\
\cmidrule(lr){3-4}\cmidrule(lr){5-6}\cmidrule(lr){7-8}\cmidrule(lr){9-10}
 &  & Q3 & Q8 & Q3 & Q8 & Q3 & Q8 & Q3 & Q8 \\
\midrule
S1D & & 73.37 & 61.46 & 75.80 & 64.27 & 70.96 & 57.41 & 73.17 & 60.53 \\
Ensemble & & 74.03 & 61.91 & 75.54 & 64.12 & 72.38 & 58.56 & 71.82 & 60.06 \\
Difference & & +0.66 & +0.45 & -0.26 & -0.15 & +1.42 & +1.15 & -1.35 & -0.47 \\
\bottomrule
\end{tabular}
\end{table}

Table \ref{tab: s1d_comp} details the $Q_{3}$ and $Q_{8}$ accuracies (expressed as percentages) for both S1D and our ensemble method. Additionally, the table displays the accuracy differences, calculated by subtracting the S1D results from those achieved by the ensemble method. The ensemble method tends to outperform the S1D model in most datasets. The most significant advantages of the ensemble method are observed in the CASP12-FM dataset, where the ensemble outperforms the S1D model by 1.42\% in $Q_{3}$ and 1.15\% in $Q_{8}$. For SPOT-2016, it outperforms S1D by 1.00\% in $Q_{3}$ and 0.59\% in $Q_{8}$; and for TEST2018, it leads by 0.66\%  in $Q_{3}$ and 0.45\% in $Q_{8}$. The S1D model, however, excels particularly in the CASP13-FM test set, surpassing the ensemble by 1.35\% in $Q_{3}$ and 0.47\% in $Q_{8}$. It also slightly outperfoms the ensemble model for NEFF1-2018 by 0.26\% in $Q_{3}$ and 0.15\% in $Q_{8}$.

\begin{figure}[h]
\centering
\includegraphics[width=0.7\textwidth]{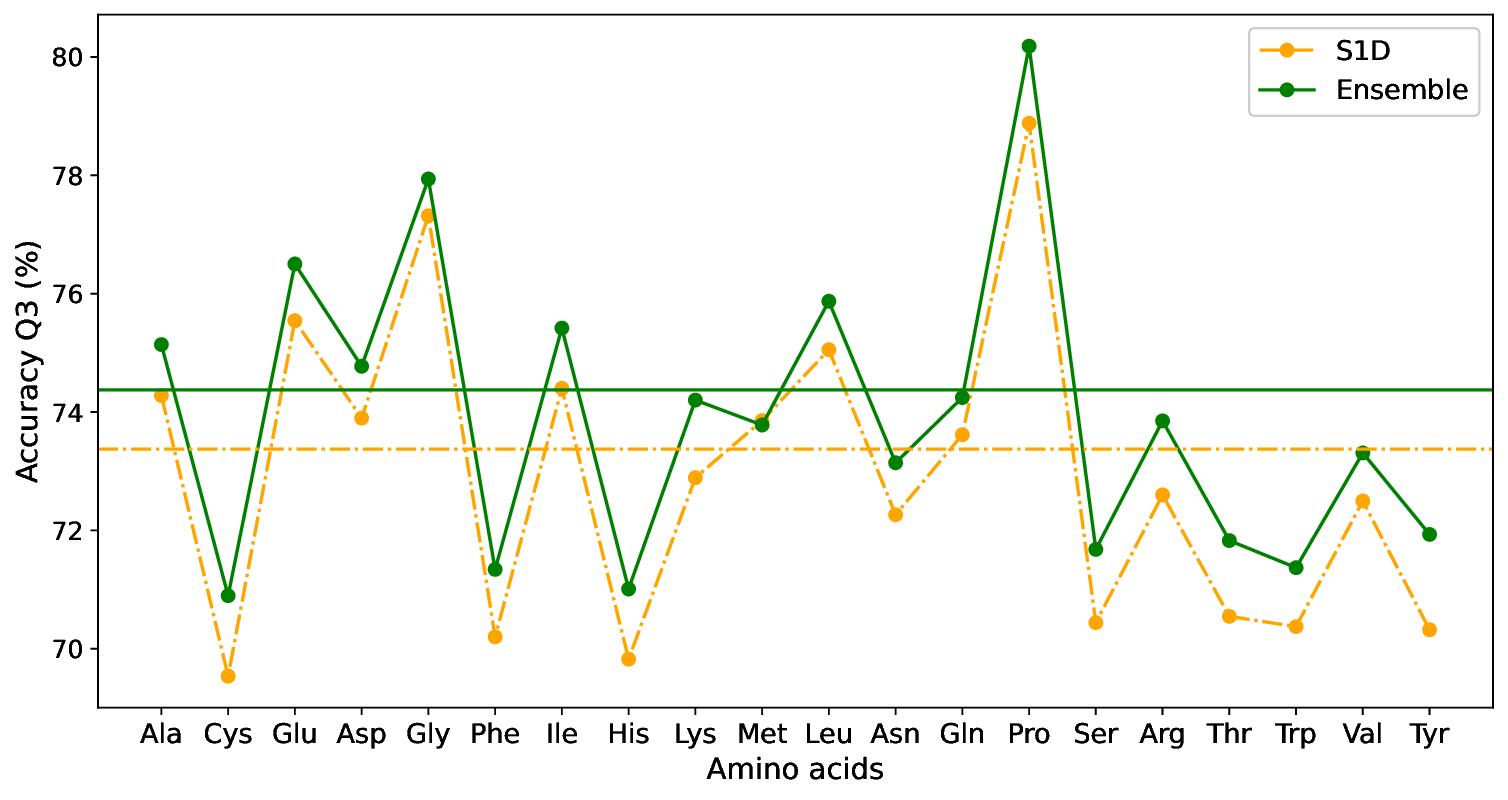}
\caption{$Q_{3}$ accuracy for individual amino acids for the ensemble model and the S1D model on SPOT-2016 dataset.}

\label{fig:aas_ensemble_s1d}
\end{figure}

Figure \ref{fig:aas_ensemble_s1d} represents $Q_{3}$ accuracy for each type of amino acid on the largest test set SPOT-2016. The straight line displays the average $Q_{3}$ accuracy across all amino acids, 73.37 for the S1D model and 74.37 for our ensemble model. The two models display a similar trend: the rare amino acids Cysteine (Cys) and Histidine (His) have the lowest $Q_{3}$ accuracy and far below the accuracy average; the abundant amino acids Proline (Pro) and Glycine (Gly) have the highest accuracy. The ensemble model consistently outperforms the S1D model for all amino acids, except for Methionine (Met), where the two models achieve similar performance.

\begin{figure}[h]
\centering
\includegraphics[width=0.7\textwidth]{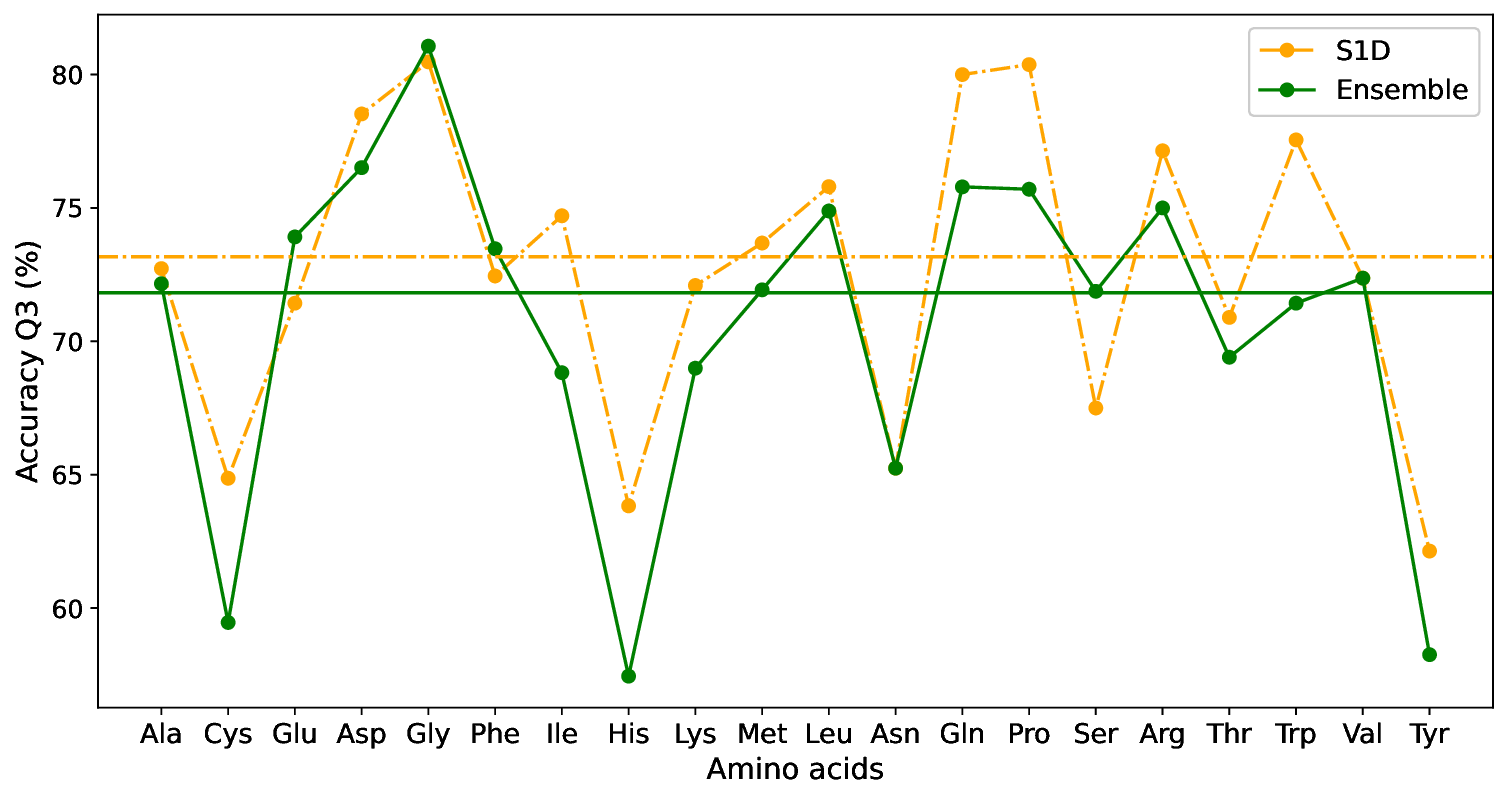}
\caption{$Q_{3}$ accuracy for individual amino acids for the S1D model and the ensemble model on CASP13-FM dataset.}

\label{fig:aas_ensemble_s1d_casp13}
\end{figure}
Figure \ref{fig:aas_ensemble_s1d_casp13} represents $Q_{3}$ accuracy for each type of amino acid on the test set CASP13-FM. Unlike the case of SPOT-2016, our ensemble model underperformed compared to S1D for 14 amino acids. It achieves similar or better performance for only six amino acids.

\begin{table}[h]
\centering
\caption{The comparison of the S1D model and the ensemble model at the sequence level.}
\label{tab:compare_s1d_ensemble_seq}


\begin{tabular}{l l c c c c c c c c} 
\toprule
Test Set & Model & \multicolumn{4}{c}{3-state} & \multicolumn{4}{c}{8-state} \\ 
\cmidrule(lr){3-6} \cmidrule(lr){7-10} 
 & & $Q_{3}$ & p-value & $n_{>}$ & $n_{=}$ & $Q_{8}$ & p-value & $n_{>}$ & $n_{=}$ \\
\midrule
SPOT-2016 & S1D & 75.33 &  & 476 &  & 62.69 & & 489 &  \\
& Ensemble & 76.43 & \(5\times 10^{-18}\) & 764  & 226 & 63.60 & \(4\times 10^{-14}\) & 723 &  254\\
\midrule
SPOT-2018 & S1D & 75.49 &  & 222 & & 61.14 &  & 220 &   \\
& Ensemble & 74.46 & \(1 \times 10^{-8}\) & 357 & 98 & 62.21 & \(5 \times 10^{-9}\) & 341 & 116\\
\midrule
TEST2018 & S1D & 76.21 &  & 81 & & 64.35 & & 103 \\
& Ensemble & 75.19 & \(5 \times 10^{-5}\) & 133 & 36  & 64.03 & 0.15 & 115 & 32\\
\bottomrule
\end{tabular}

\bigskip
\raggedright 
Note: $n_{>}$, the number of sequences for which one model achieves higher $Q_{3}$ or $Q_{8}$ accuracy than the other model; $n_{=}$, the number of sequences for which both models achieve equal $Q_{3}$ or $Q_{8}$ accuracy.
\end{table}

Table \ref{tab:compare_s1d_ensemble_seq} presents the $Q_3$ and $Q_8$ accuracies at the sequence level for the test sets where the performance difference between the S1D model and our ensemble model is statistically significant. The two models exhibit statistically significant differences for SPOT-2016 and SPOT-2018, as indicated by p-values significantly lower than 0.05. These p-values were obtained using the Wilcoxon signed-rank test\cite{wilcoxon1945individual}. Notably, our ensemble model outperformed the S1D model for the majority of sequences. Notably, there are instances where both models achieved identical accuracies. For TEST2018, the models showed statistical significance in $Q_3$ accuracy (p-value $=$ \(5 \times 10^{-5}\) $ < 0.05$) but not in $Q_8$ accuracy (p-value $= 0.15$  $ > 0.05$ ).

The OHE, AFE, and MFE models each require approximately 3.9 million trainable parameters. The S1D model necessitates over 35 million trainable parameters, which is about nine times more than each of our encoding models. Despite utilizing significantly fewer trainable parameters, our ensemble model surpasses the S1D model in performance across most test sets, including the challenging CASP12-FM dataset. This efficiency enables us to train each encoding model on cost-effective GPUs with limited memory.

\subsection{Sequence length dependence}
At the residue level, the accuracy differences between each pair of encoding methods are on a similar level as discussed in Session \ref{sec:ensemble_performance}. At the sequence level, the Wilcoxon statistical test for the SPOT-2016 dataset shows that the differences in $Q_{3}$ accuracies obtained using MFE and OHE models, and MFE and AFE models, are statistically significant, as indicated by p-values of 0.0007 and 0.004, respectively. However, the performance difference between AFE and OHE models is not statistically significant (p-value $= 0.876 > 0.05$).

\begin{figure}[h]
\centering
\includegraphics[width=\textwidth]{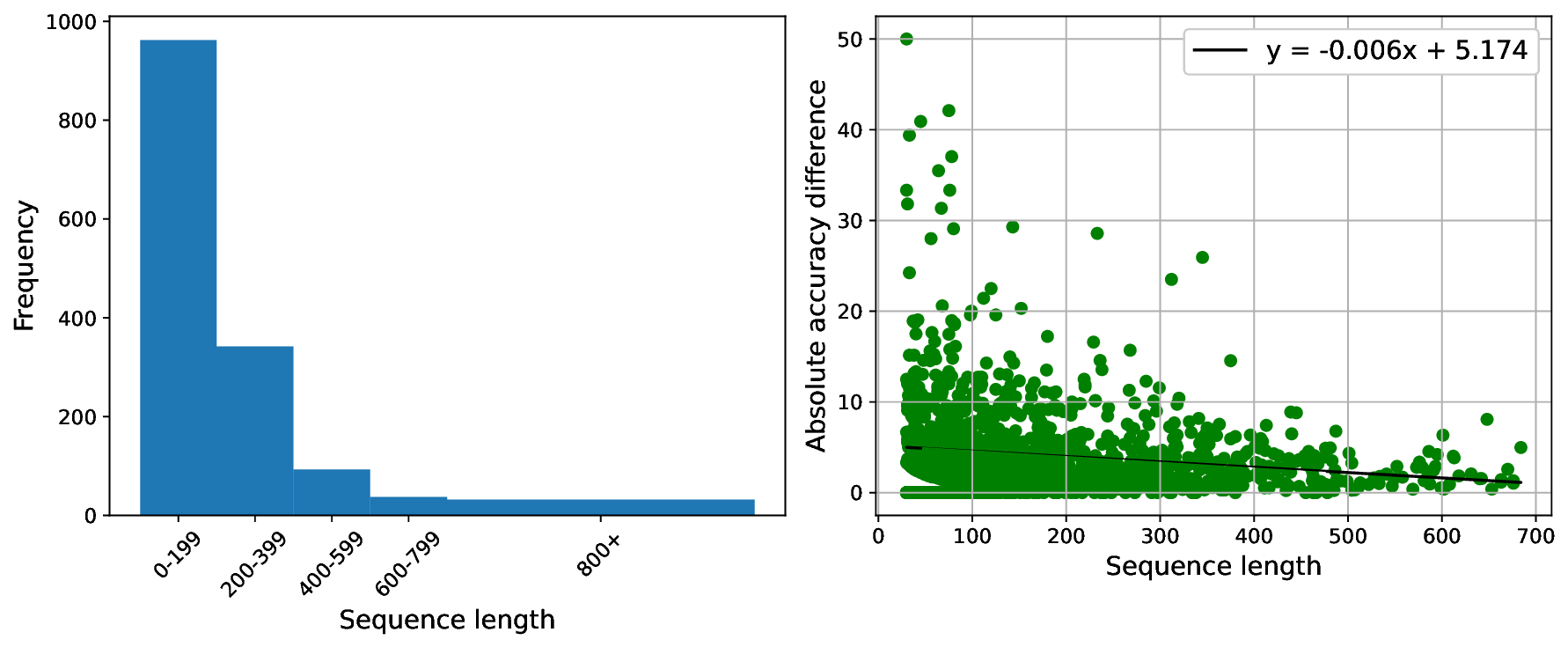}
\caption{The performance difference of OHE and AFE models at the sequence level for the SPOT-2016 dataset. Left panel: Distribution of sequence lengths. Right panel: Dependence of the absolute accuracy difference between the models on sequence length. The black line represents the linear regression fit, with a slight negative correlation ($y = -0.006x + 5.174$).}

\label{fig:linear_reg}
\end{figure}

Figure \ref{fig:linear_reg} examines the absolute $Q_{3}$ accuracy differences between OHE and MFE models for the SPOT-2016 dataset at the sequence level. The left panel demonstrates the distribution of sequence lengths. The histogram shows that most sequences fall within the shorter length categories (0-199) and (200-399), with a significant decrease in sequence count as length increases. The right panel presents the relationship between the absolute accuracy difference of the two models and sequence length. The black line represents the least square regression fit, indicating a slight negative correlation ($y = -0.006x + 5.174$) between sequence length and absolute accuracy difference. The linear regression fit is confined to sequences shorter than 700 residues due to the sparse representation of longer sequences in the dataset. This sequence length limit has been used in previous studies\cite{mufold, len_limit}. As sequence length increases, the absolute accuracy difference decreases and approaches zero at a sequence length of 700 residues. This observation suggests that the performance difference in $Q_{3}$ accuracy between OHE and MFE models is most significant for shorter sequences.

The linear relationship between absolute accuracy differences and sequence length is statistically significant, indicating by F-statistic\cite{fisher1925statistical} of 36.27 and p-value\cite{fisher1925statistical} of \(2.18 \times 10^{-9}\). However, the distribution of the residuals in the linear regression fit is highly skewed and could affect the validation of the F-test. To confirm that this relationship is not due to chance, we conducted a quantile linear regression\cite{koenker2005quantile}, which does not require normally distributed residuals for F-test validity. Quantile regression models the conditional quantile of the dependent variable $Y$, in this case, absolute accuracy difference, given the independent variable $X$, the sequence length, as $Q_{Y}(\tau|X) = X\beta(\tau)$. In this expression,  $\tau$ represents the quantile, and $\beta(\tau)$ represents the coefficients. This method allows for modeling various quantiles, such as the median ($\tau = 0.5$) or the 25th and 75th percentiles, offering a comprehensive view of how sequence length affects accuracy differences across the distribution.

\begin{figure}[h]
\centering
\includegraphics[width=\textwidth]{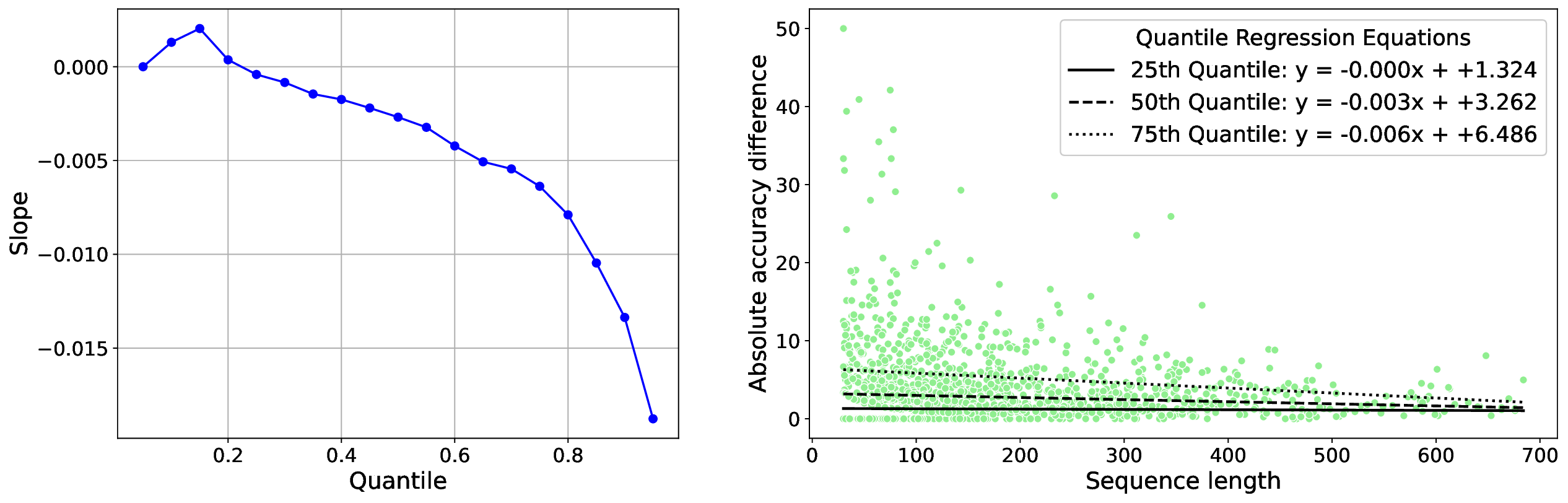}
\caption{The relationship between the absolute accuracy difference of OHE and MFE models and sequence length across quantiles. Left panel: Coefficient of sequence length across quantiles. Right panel: Quantile regression lines at different quantiles. $y$ represents the conditional quantile, and $x$ represents the sequence length.}

\label{fig:quantile_reg}
\end{figure}

Figure \ref{fig:quantile_reg} left panel shows the effect of sequence length on absolute accuracy difference across quantiles, and the right panel demonstrates the quantile regression at $25^{th}, 50^{th}, 75^{th}$ quantile. As the quantile increases, the coefficient of sequence length becomes more negative. At lower quantiles, the effect of sequence length on accuracy difference is negligible, suggesting limited influence at smaller absolute accuracy differences. Conversely, the coefficients are increasingly negative at higher quantiles, particularly beyond the median. This highlights a strong, negative impact of increasing sequence length on larger initial differences.

The observations from both least squares and quantile regression analyses suggest that longer sequences are more consistently predicted by the OHE and MFE models. An important observation is that the 3-state predictions by these two models exhibit the most significant differences for shorter sequences. For future work, we plan to focus on leveraging the diversity in ensemble predictions to increase $Q_{3}/Q_{8}$ accuracy specifically for shorter sequences.

\section{CONCLUSIONS}

In this study, we developed two new encodings for each type of amino acid based on Morgan and atom-pair fingerprints. Leveraging the FastMap algorithm, we transformed these high-dimensional fingerprints into dense vectors with dimensions of just 18 and 14. The new encodings are as effective as one-hot encoding, demonstrating the versatility of molecular fingerprints not only for drug design but also for protein structure prediction. Notably, the accuracy of the one-hot encoding model improved when combined with the predictions from the chemical encoding models. This suggests that chemical encodings provide additional information that a limited-size LSTM-based model cannot learn from one-hot encoding alone. The ensemble of one-hot and chemical encodings enhances the accuracy of the one-hot encoding model for most amino acid types, proving the effectiveness of the FastMap algorithm in preserving chemical distances between amino acids and improving the ensemble model's ability to capture their interactions within protein sequences. Our ensemble of one-hot and chemical encoding models is competitive with the LSTM-based model of SPOT-1D-Single. Each model in our ensemble requires nine times fewer trainable parameters than the S1D model, facilitating training with larger datasets or limited computational resources. While we utilized LSTM-based models in this study, we anticipate that other deep-learning models could benefit from the ensemble of one-hot and chemical encoding.

We limited our work to studying protein secondary structure, but we expect the new chemical encodings to have applications in predicting other protein properties, such as dihedral and torsion angles. Although the FastMap algorithm preserves the chemical similarity/dissimilarity between amino acid types, it still loses some information on amino acid structures. With the advancement of graph neural networks, it would be interesting to use amino acid structures directly as inputs. This might offer advantages over fragmenting molecule structures like the Morgan and atom-pair fingerprint methods.

\subsection*{AUTHOR CONTRIBUTIONS}
\textbf{Hoa Trinh} identified the research question, developed methodologies, implemented all the algorithms, conducted all the experiments, analyzed all the data, and wrote the manuscript. \textbf{T. K. Satish Kumar} suggested the use of the FastMapSVM framework, molecular fingerprints, and chemical distance functions. 

\subsection*{ACKNOWLEDGMENTS}
We thank Prof. Hoang Trinh for his insightful feedback on the manuscript draft. We thank Ang Li for sharing his FastMap Support Vector Machine code. Additionally, we acknowledge the use of data from the SPOT-1D-Single study.

\end{document}